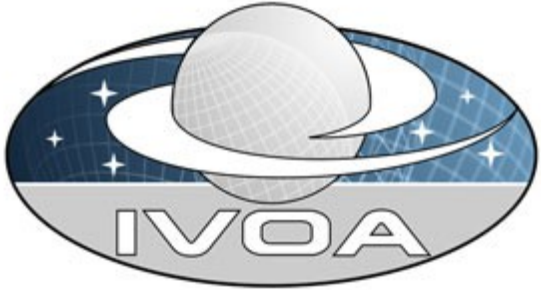

**I**nternational

**V**irtual

**O**bservatory

**A**lliance

# MOC – HEALPix Multi-Order Coverage map Version 1.0

*IVOA Recommendation 2 June 2014*

**This version:**
　　1.0: Recommendation 2014-06-02

**Previous version(s):**
　　None

**Interest/Working Group:**
　　Applications: http://www.ivoa.net/twiki/bin/view/IVOA/IvoaApplications

**Editor:**
　　Pierre Fernique

**Authors:**
　　Thomas Boch, Tom Donaldson, Daniel Durand, Pierre Fernique, Wil O'Mullane, Martin Reinecke, Mark Taylor

## Abstract

This document describes the Multi-Order Coverage map method (MOC) to specify arbitrary sky regions. The goal is to be able to provide a very fast comparison mechanism between coverage maps. The mechanism is based on the HEALPix sky tessellation algorithm. It is essentially a simple way to map regions of the sky into hierarchically grouped predefined cells.

## Status of This Document

This document has been reviewed by IVOA Members and other interested parties, and has been endorsed by the IVOA Executive Committee as an IVOA Recommendation. It is a stable document and may be used as reference material or cited as a normative reference from another document. IVOA's role in making the Recommendation is to draw attention to the specification and to promote its widespread deployment. This enhances the functionality and interoperability inside the Astronomical Community.

A list of [current IVOA Recommendations and other technical documents](http://www.ivoa.net/Documents/) can be found at http://www.ivoa.net/Documents/.

## Contents

This document has three sections: the first section is descriptive and provides the MOC idea, justification, usage, and examples; the second section is normative and describes the MOC coding. Additional information concerning MOC programming technique is discussed in the third section and in the appendixes.

# 1 Introduction

The encoding method described in this document allows one to define and manipulate any region of the sky in such a way that basic operations like union, intersection, equality test can be performed very efficiently. It is dedicated to VO applications or VO data servers for building efficient procedures for which mapping knowledge is required like generic catalog cross-match, computation of data set intersections, or similar operations.

This encoding method is called "Multi-Order Coverage map" or "MOC".

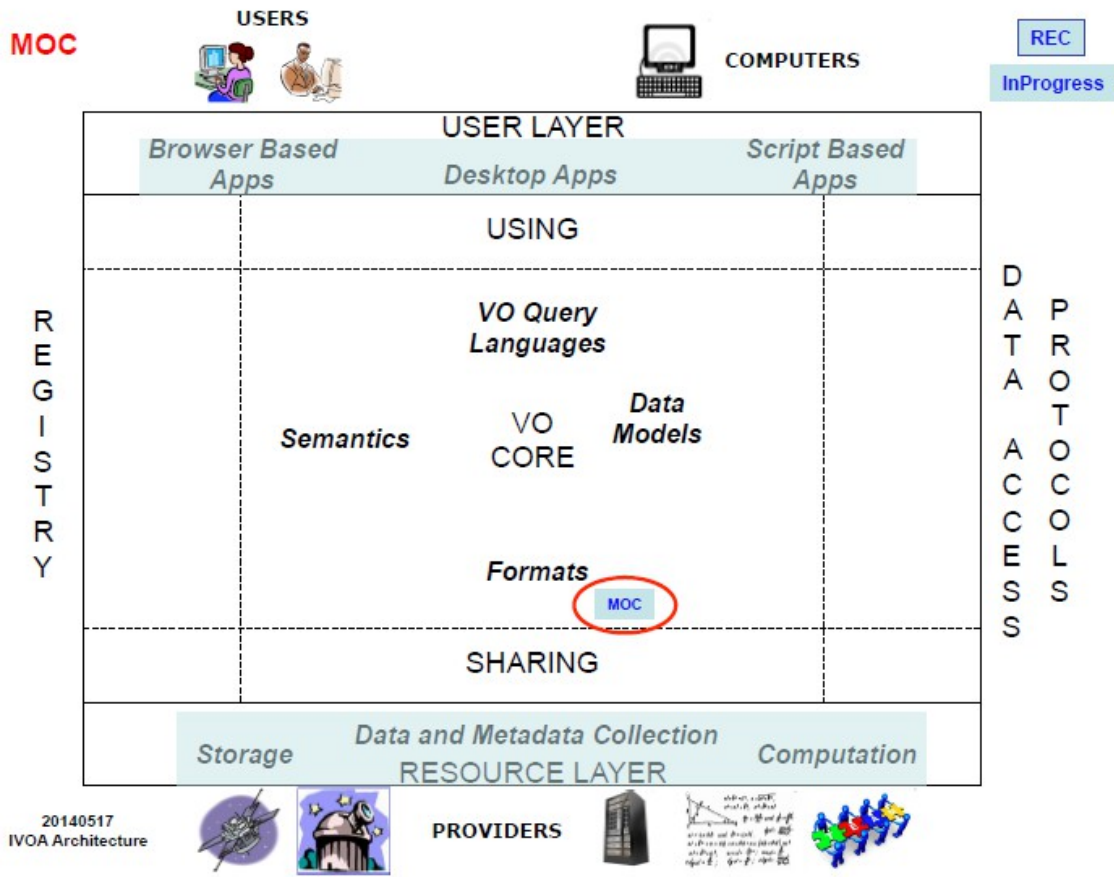

The MOC standard describes a standalone format and is neither a dependent, nor a dependency, of other VO standards. The figure above illustrates MOC in the context of the overall IVOA Architecture.

## 1.1 The idea

The goal behind the MOC is to get a spherical geometry description for any sky region in order to provide very fast and accurate union, intersection and equality operations between them. In order to accomplish this task, we based the system on a regular and hierarchical partitioning of the sphere, a process called tessellation.

The tessellation technique we used is called HEALPix [1] which divides the sphere into 12 diamonds, each of them sub-divided into 4 diamonds recursively. Thus the sphere at level 1 will consist of 48 diamonds, 192 diamonds at level 2, 768 at level 3 and so on where each diamond at a given level is covering an equal area of the sphere.

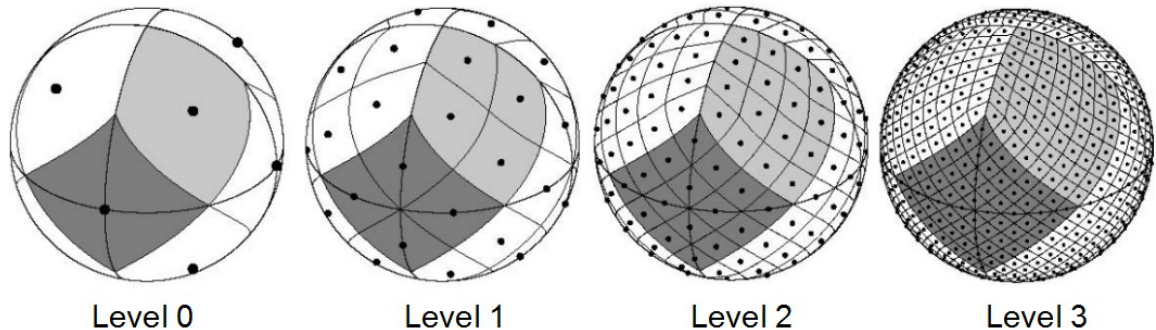

So any region of the sky can be defined by the list of the diamond indices involved to define the region. Union, intersection and equality operations on such lists are straightforward and very fast - even more so if this list has been previously sorted.

In order to reduce the volume of these lists any four consecutive diamonds are replaced by their "parent", and this, recursively. The deepest level is set by the resolution required for describing the data set.

## 1.2 Example

The HEALPix diamonds covering the blue sky region drawn in the figure below corresponds to the diamond indices 1164 to 1215, 1226, 1536 to 1539, 5628 to 5631 and 5973, at level 5 (cf. numbering scheme in 2.2.2). According to the hierarchical HEALPix principle, this list can be also expressed by the 3 shorter lists: level 3: 73,74,75, level 4: 291,384,1407 and level 5: 1226,5973. These three lists, alternatively represented as eight (level,index) pairs, constitute the level 5 MOC containing this sky region.

In more detail: the pixels are positioned on the celestial sphere using the HEALPix "equatorial" convention (section 2.2.1), pixel indices for each level are assigned using the HEALPix NESTED convention (section 2.2.2), and each (level,index) pair is packed into a single integer using the NUNIQ encoding described in section 2.3.1. A MOC can thus be represented as a flat list of integers (in this example, 8 of them) and stored in a single-column FITS table (section 2.3.2).

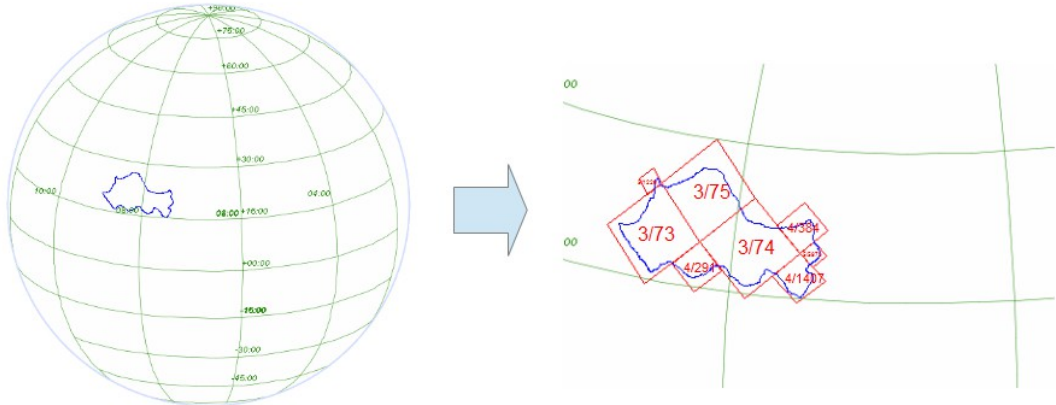

```
3: 73,74,75
4: 291,384,1407
5: 1226,5973
```

## 1.3  Illustrations

- This illustration represents an excerpt of the GLIMPSE survey and its corresponding MOC:

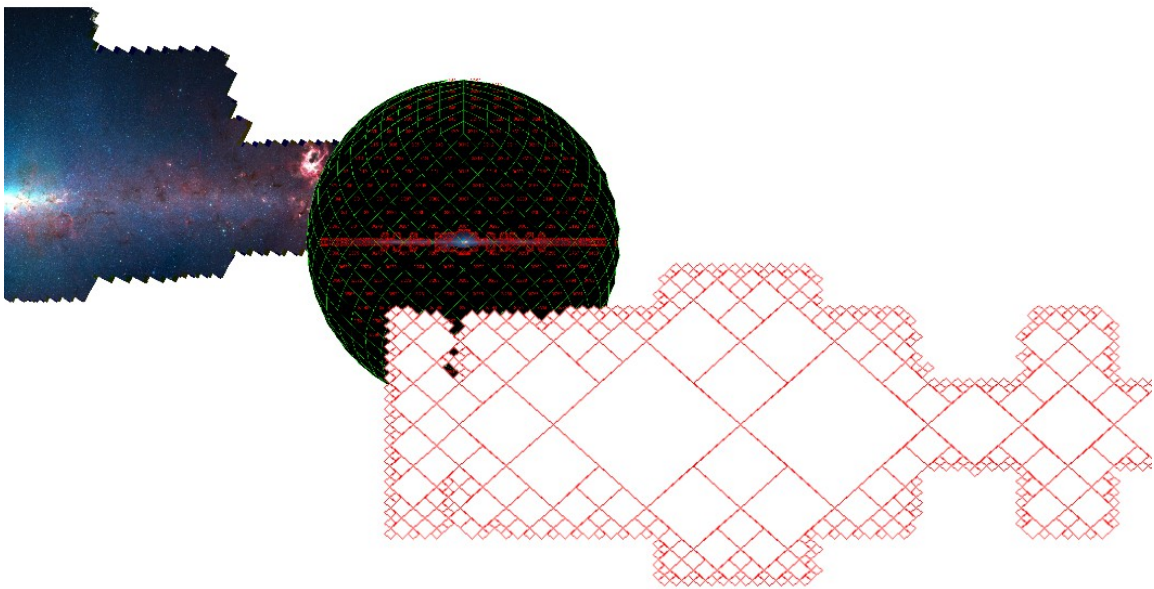

- Two MOCs corresponding to GALEX AIS (HEALPix level 8 – 70,000 cells) and SDSS (HEALPix level 10,- 225,000 cells):

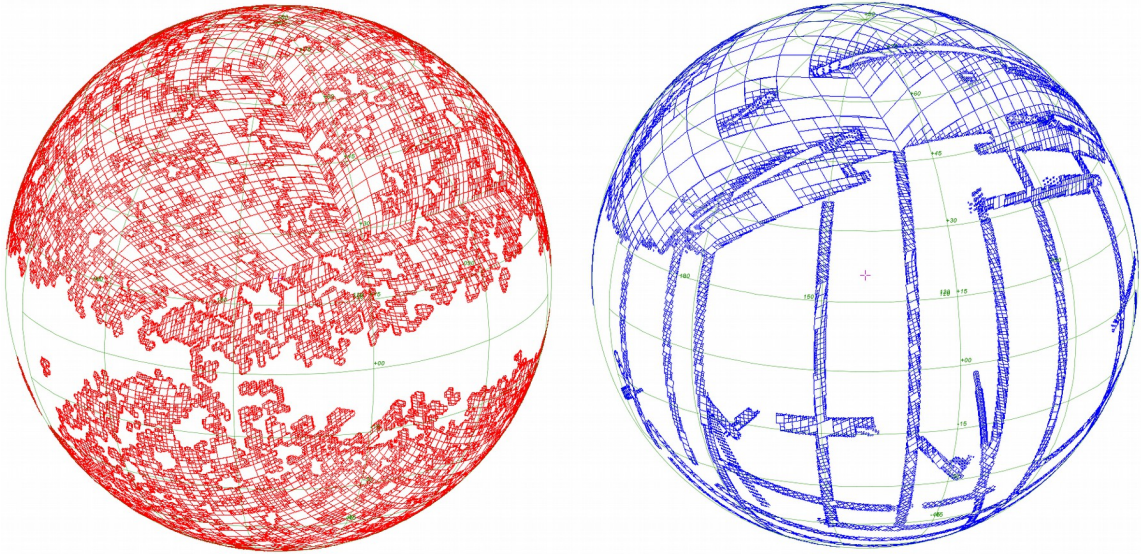

## 1.4 HEALPix tessellation

Defining a sky region by a subset of regular sky tessellation or tiles is not a new idea. The advantage of sharing a homogeneous tessellation method between different data centres has already been described in several articles, notably in the VO context [2], [3].

In astronomy, one could find three main methods of partitioning the sphere : Q3C, HTM and HEALPix which are respectively using squares, triangles and diamonds for mapping the sky.

Several articles have compared these methods [4]. We justified the choice of HEALPix for the MOC because of these these four points:
- **Equal areas:** by construction, HEALPix consists of diamonds with equal spherical surfaces. Thus the area of a given region is trivial to compute;
- **Computing time:** HEALPix has the peculiarity that the computing time does not depend on the hierarchical level (no recursive algorithm);
- **Accuracy:** HEALPix provides libraries which allow the calculation up to accuracy of 0.4 mas (level 29) [8];
- **Standard:** Existence of many HEALPix libraries: C++, Java, Fortran, IDL... Also, there is an already defined standard for encoding the HEALPix maps into a simple FITS file. In addition, HEALPix was selected for several all sky missions such as WMAP, Planck and GAIA. The HEALPix main web site is located at Jet Propulsion Lab [5].

## 1.5 Usage examples

### 1.5.1 Coverage visualization

The MOC structure allows easy visualization of any data set coverage. The MOC meshing nature provides an easy interpretation. When using a simple 4 diamond corner drawing approximation as shown in figure section 1.3, the time required for drawing this complex coverage is only a few milliseconds. If necessary, because of the MOC's hierarchical structure, this time can be reduced by stopping the drawing process at a coarser level than the best resolution available.

### 1.5.2 Comparing the coverage of multiple data sets

The computation of data set intersections using the MOCs is simple (it is simply a list comparison). The result of any operations is itself a MOC which can be used in further operations. For instance it is possible to compute the sky region determined by the intersection of two MOC surveys, and subsequently, query a database for retrieving objects whose position falls within this intersection.

### 1.5.3 Query databases using MOC

In principle, querying a positional database using complex sky region coverage is possible using ADQL[1]. However, this is rarely possible when the sky region has to be described as unions and intersections of sub-regions to cover a complex, non regular area. In practice, most existing ADQL implementations are only supporting simple regions (cones, boxes, polygons), and can rarely deal with unions and intersections unless by joining independent queries – and this even if the described region is, in fact, empty!

If databases are adapted to supporting MOC based queries, they will offer then a useful method allowing any kind of sky region query. In addition, if the internal spatial index of the database is itself based on HEALPix, the filtering will then be straightforward and all the intermediate sky computations will be removed providing an optimal response time.

### 1.5.4 Multi-site positional search

Often, a typical query from a virtual observatory (VO) user is to request all possible records from the VO at a given sky position. While this is in principle possible by dispatching one narrow positional query to every registered Cone Search, SSA or SIA service, in practice, the number of queries required leads to an unacceptable load on both clients and services. Moreover, most of these queries will deliver no results since most services often lack coverage in the queried region. If basic footprint/coverage information was available for all registered services, only those with coverage in the region of interest would then be queried. This would provide a great reduction in the number of services to be queried optimizing the response time. Using the MOC offers the opportunity to provide this coverage information in a uniform way. The MOC could be stored locally for a given service or centrally where the coverage for a number of services would be supported.

---

[1]Astronomical Data Query Language (ADQL) [6] is a recommended IVOA standard based on SQL and integrating geometrical functions (see also STC standard [7]).

## 1.6 Scope and performance

### 1.6.1 Scope of usage

MOC is designed for defining coverage regions of data sets, from pixel surveys to source catalogs.

Even though MOC allows to describe regions on the sky, it is not possible to exactly define these regions. If this is required, it is better to use a region definition based on STC[2]. For instance, MOC is clearly not suitable to define an instrument field of view on the sky. In other words, simple geometric objects such as circles and polygons are better defined using an STC region than using the MOC and this is true when an **exact** mapping is required.

But for comparing two regions, it is definitely easier and faster using MOC rather than STC. The main STC problem when dealing with region comparison is that there is no canonical way of expressing a region, ie, various STC literal expressions can specify the same region. Thus, a basic equality test with STC ends up being a complex spherical geometry problem. And solving it with software can be a really difficult task. But a MOC region is always described in one way – a list of cells – so the comparison algorithms are simple and fast.

MOC is a coverage map, not a density map. Concretely, there is no value associated to sky cells. This choice allows to reduce the size of MOC "by factorizing adjacent cells". Notice that a standard HEALPix map can be used as a density map, and can be fully comparable with MOC (see below MOC encoding section).

---

[2]STC is a recommended IVOA standard which allows to define a sky region by operations (union, intersection...) with simple shapes (circles, ellipses, polygons...) [7]

### 1.6.2 MOC resolution

The MOC resolution is set by the maximum value of the HEALPix level used to define a region. Its selection depends on the accuracy chosen by the provider to define the region.

As data set boundaries are not aligned with the HEALPix cell borders, a MOC is generally an upper-approximation of the data set coverage. The quality of this approximation depends directly on the chosen MOC resolution.

**Note**: There is no problem to compare MOCs with different resolutions.

The following table provides the HEALPix cell angular resolution for each HEALPix level.

| Level | Best cell resolution(*) | Level | Best cell resolution(*) |
|---|---|---|---|
| 0 | 58.63° | 15 | 6.442" |
| 1 | 29.32° | 16 | 3.221" |
| 2 | 14.66° | 17 | 1.61" |
| 3 | 7.329° | 18 | 805.2mas |
| 4 | 3.665° | 19 | 402.6mas |
| 5 | 1.832° | 20 | 201.3mas |
| 6 | 54.97' | 21 | 100.6mas |
| 7 | 27.48' | 22 | 50.32mas |
| 8 | 13.74' | 23 | 25.16mas |
| 9 | 6.871' | 24 | 12.58mas |
| 10 | 3.435' | 25 | 6.291mas |
| 11 | 1.718' | 26 | 3.145mas |
| 12 | 51.53" | 27 | 1.573mas |
| 13 | 25.77" | 28 | 786.3µas |
| 14 | 12.88" | 29 | 393.2µas |

*(*) mean value. The HEALPix cell has constant area, not constant linear dimensions.*

### 1.6.3 Size of MOC

The MOC describes a sky region as an explicit list of cells. The resulting size of this list can vary a lot from a few bytes to several kilobytes and more. Since MOC is hierarchical, its size mainly depends on two factors:

1. The chosen MOC resolution;

2. The geometry of the region ("compactness").

When MOCs are used to describe pixel survey coverages, the number of cells depends on the size of the coverage perimeter (size of borders).

When we want to describe source catalog coverages, the number of cells depends of the distribution and the density of the sources in the catalog (for instance, for a catalog with a low density and a high distribution, it is possible to choose a very good MOC resolution without increasing the MOC size significantly – see appendix B).

### 1.6.4 MOC performance

The required MOC operations are identical to classical set operations: equality, union, intersection... The algorithms implementing these operations are based on simple list manipulations. Good performance can be obtained in using dichotomic algorithms on sorted lists. Typically, the computation of intersection of two complex MOCs – (for instance using GALEX and SDSS MOCs as illustrated above) took only a few milliseconds to compute. The proposed algorithms are in appendix A.

## 2 MOC encoding

### 2.1 Vocabulary

The keywords "must", "required", "should", and "may" used in this document are to be interpreted as described in the W3C specifications (RFC 2119 [9]). Mandatory elements are indicated as **must**, recommended elements as **should**, and optional elements as **may** or simply "may" without the bold face font.

As HEALPix is already used for other purposes, it has its own defined vocabulary. A HEALPix level is called an **ORDER** and a cell index is called a **NPIX**.

### 2.2 MOC particularities

#### 2.2.1 Coordinate system

HEALPix allows three coordinate systems: galactic, equatorial and ecliptic. In our implementation, allowing various coordinate systems would limit the possibility to compare efficiently MOCs. There is indeed no equivalence between an HEALPix cell described in a given coordinate system and a cell, or a list of sub-cells expressed in a different coordinate system. Consequently, the MOC definition **must** be expressed in equatorial coordinate using the ICRS reference. This choice has been motivated by looking at most catalogs and realizing that most of them are using equatorial coordinates. Adopting this restriction has no impact on specialized MOC usage describing whole sky surveys such as WMAP and PLANCK.

### 2.2.2 NESTED numbering

Another adopted restriction concerns the numbering scheme used in MOC for specifying the cell indices. HEALPix supports both "NESTED" and "RING" numbering schemes [1]. Though it is possible to convert the cell indices from one system to the other, it takes a significant amount of time. In order to follow the guidelines trying to implement fast comparisons, MOC **must** use the NESTED numbering scheme only.

The NESTED HEALPIX numbering consists of enumerating all cells in a specific order. For instance, at level 1, there are 48 diamonds (12x4) enumerated from 0 to 47. In the NESTED scheme, the 4 sub- diamonds of diamond M have the indices: (Mx4)+3, (Mx4)+1, (Mx4)+2, (Mx4) in reading order. And reciprocally, the parent index of diamond N is N/4.

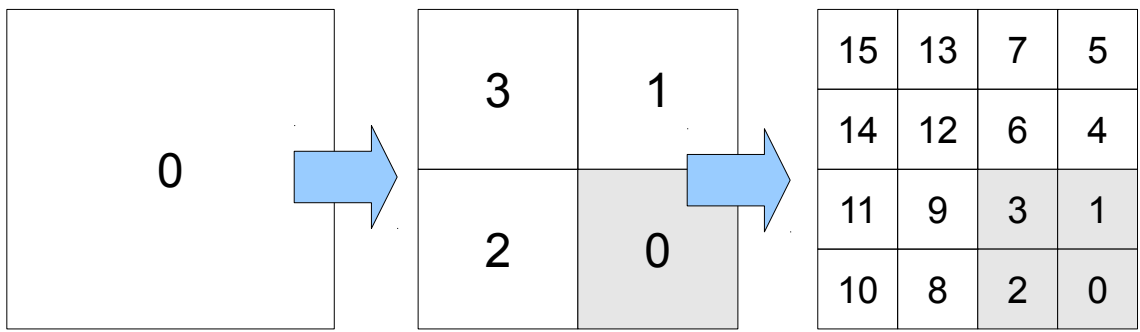

**Note:** The level 0 is a special case, it contains only 12 diamonds enumerated from 0 to 11.

### 2.2.3 Well-formed MOC encoding

An encoded MOC **must** also be "well-formed". To keep the canonical property of MOC, the redundant cells are not allowed and the hierarchical encoding principle **must** be respected. Thus it is not allowed to encode 4 sibling cells instead of their parent (only at level 0 for the 12 first diamonds). Using this simple constraint, there is one and only one way to describe a specific region and this makes the interoperability principle easy to implement.

### 2.2.4 Sky coordinates and HEALPix cells

The mechanism used to determine which HEALPix cell contains a given sky location is described in the main article defining the HEALPix system [1]. Several support libraries supporting the most important set of primitives are already available [8].

These libraries are required if one wants to generate MOCs, and also wants to compare them with sky coordinates (see 3.5). Though please note that these libraries do not have built-in support to performing basic MOC arithmetics.

## 2.3 MOC serialization

### 2.3.1 NUNIQ packing

Each MOC cell is defined by two numbers: the hierarchy level (HEALPix ORDER) and the pixel index (HEALPix NPIX). The NUNIQ scheme[3] defines an algorithm for packing an (ORDER,NPIX) pair into a single integer for compactness:

```
uniq = 4 * (4^order) + npix
```

The inverse operation is:

```
  order = log2(uniq/4) /2
npix = uniq – 4 * (4^order)
```

For ORDER<=13 these UNIQ values can be stored in a 32-bit signed integer, and for ORDER<=29 in a 64-bit signed integer.

### 2.3.2 FITS serialization

To encode a MOC in a FITS file, each cell is converted to a UNIQ integer using the NUNIQ scheme, and the resulting list is stored in a single-column binary table extension. The list of cells **must** be well-formed, and the UNIQ values **must** be in ascending numerical order.

The single column **must** have an integer type, which **should** be 32-bit (TFORM1='J') for a MOC resolution <=13 or 64-bit (TFORM1='K') for higher resolutions.

The rest of this section describes the FITS headers used to annotate this table.

*Example of FITS headers for a MOC:*

```
SIMPLE  =                    T
BITPIX  =                    8
NAXIS   =                    0
EXTEND  =                    T
END

XTENSION= 'BINTABLE'           / HEALPix Multi Order Coverage map
BITPIX  =                    8
NAXIS   =                    2
NAXIS1  =                    4
NAXIS2  =                16461
PCOUNT  =                    0
```

---

[3] Eric Hivon proposal (HEALPix co-author [1])

```
GCOUNT   =                    1
TFIELDS  =                    1
TFORM1   = '1J      '
TTYPE1   = 'UNIQ    '           / HEALPix UNIQ pixel number
PIXTYPE  = 'HEALPIX '           / HEALPix magic code
ORDERING = 'NUNIQ   '           / NUNIQ coding method
COORDSYS = 'C       '           / ICRS reference frame
MOCORDER =                   12 / MOC resolution (best order)
MOCTOOL  = 'Aladin7.5 '         / Name of the MOC generator
MOCTYPE  = 'CATALOG'            / Source type (IMAGE or CATALOG)
MOCID    = 'ivo://CDS/I/259'    / Identifier of the collection
ORIGIN   = 'ivo://CDS'          / MOC origin
DATE     = '2013-06-15T11:50:43' / MOC creation date
EXTNAME  = 'Tycho MOC'          / MOC name
END
```

**HEALPix mandatory keywords:**

As the FITS MOC is derived from a more generic standard dedicated to HEALPix maps, the FITS header **must** contain the following keywords/value.

```
PIXTYPE  = 'HEALPIX '           / HEALPix magic code
ORDERING = 'NUNIQ   '           / NUNIQ coding method
COORDSYS = 'C       '           / ICRS reference frame
```

**MOC mandatory keywords:**

The MOC **must** have the keyword:

- **MOCORDER:** Is the best resolution of the MOC file expressed as a HEALPix order.

**MOC optional keywords:**

The following keywords **may** be provided to provide additional MOC-specific information:

- **MOCTOOL:** Is the name of the MOC generator software. It is also recommended to add the software version number to its name.

- **MOCTYPE**: Indicates the input type used to produce the MOC. We have two different types. The first one, 'IMAGE' includes any areas computed from existing images and/or footprint or even STC strings. The second type, 'CATALOG' indicates that the MOC was computed from a collection of point sources using a unique and/or derived area.

- **MOCID**: Specify an identifier for the original data used to build the MOC. We suggest using IVOA identification like "ivo://CADC/HST/HLA/WFPC2/F555W' describing the HST HLA archive of the WFPC2 instrument using the F555W filter. If required, the provenance of the original data may be fully described using the classical FITS vocabulary: PROJECT, INSTRUME, FILTER, VERSION...

**FITS keywords in MOC context:**

It is recommended to include the following standard keywords in accordance with section 4.4.2 of the FITS standard:

- **ORIGIN**: This keyword could be use to store the name or the VO identifier of the MOC origin. We suggest using IVOA identification like ivo://CADC, ivo://IPAC.
- **DATE**: Could be the creation/generation date of this MOC file (ISO date format)
- **EXTNAME**: This standard FITS keyword could be used as a short label/name for the MOC. It would then be easier to distinguish the different MOCs when displayed within the same tool when doing overlay.

# 3 Help for implementing

## *3.1 ASCII MOC*

In general the FITS encoding described in section 2 should be used for exchange of MOCs. However, if it is required to write a MOC as an ASCII string (for a web form, for debugging, ...) it is suggested to use one of the following syntaxes:

### 3.1.1 JSON syntax

A JSON MOC **may** be written following this syntax:
{ "order":[npix,npix,...], "order":[npix, npix...], ... }.

Example of a JSON MOC
```
{"1":[1,2,4], "2":[12,13,14,21,23,25]}
```

### 3.1.2 ASCII string syntax

An ASCII string MOC **may** be written following this syntax:
order/npix,npix,... order/npix,npix.

The usage of a range operator is allowed in the list of npix using the dash ("-") as a separator: lownpix-hightnpix.
Warning: In this basic simple ASCII string format only the values **may** be not sorted, and the MOC **may** be not well-formed.

Example of a ASCII string MOC
```
1/1,3,4 2/4,25,12-14,21
```

## 3.2 Basic HEALPix functions

For generating MOC from observations, or drawing them on the sphere, an HEALPix library is required. The basic functions available in all HEALPix libraries are the following :

- *npix* <= **coordToNpix**(*order, alpha,delta*) : returns the HEALPix cell index containing the alpha,delta coordinates.

- *ArrayOfNpix* <= **queryDisc**(*order, alpha,delta,radius*) : returns the list of cell indices covering the (long,lat,radius) cone

- *ArrayOfNpix* <= **queryPolygon**(*order, alpha1,delta1, … alphaN,deltaN*): returns the list of cell indices covering the spherical polygon

- *(alpha,delta)* <= **NpixToCoord**(*order,npix*) : returns the coordinates of the center of order/npix cell.

- *ArrayOf(alpha,delta)* <= **NpixToCorners**(*order,npix*) : returns the corner coordinates of order/npix cell.

# Appendix A – Suggested algorithms for basic operations

Mapping a MOC to a unique sorted list of cells at the deepest resolution level allows usage of very easy and very fast algorithms. Basic operations such as unions or intersections can be computed via bit shifts and simple dichotomic algorithms on sorted lists.
To reduce as much as possible the memory requirement, a good practice is to store range sets of continuous cells [minValue .. maxValue[, instead of individual cells.

## *Union: moc1 ∪ moc2*

    **Map** moc1 to rangeList
    **Map** moc2 in the same rangeList
    **Unmap** the resulting rangeList

## *Intersection: moc1 ∩ moc2*

    **Map** moc1 in rangeList1
    **foreach** order/npix **of** moc2
        shift=2*(maxOrder-order)
        **append** in a rangeList2 **the intersection between**
            **[**npix << shift .. npix+1 << shift**[**
        **and** the corresponding range(s) of rangeList1
    **Unmap** rangeList2

## *Map: moc To rangeList*

    **foreach** order/npix **of** moc
      shift=2*(maxOrder-order)
      **append in rangeList [**npix << shift , (npix+1) << shift**[**
      *(the range overlapping must be adjusted)*

## *Unmap: rangeList To moc*

    **for** order = 0 to maxOrder
        **end if** rangeList is empty
        shift = 2*(maxOrder-order)
        offset = (1<<shift) -1
        **foreach** range **[**min..max**[** of rangeList
            **append** in moc order/npix **where** npix is in **[**m1 .. m2**[**
                m1 = (min+offset) >> shift
                m2 = max >> shift
            **remove from rangeList [**m1<<shift .. m2<<shift**[**

## Appendix B – MOC size vs MOC resolution

The size of MOC for sources catalogs depends on the density and the distribution of these sources over the sky. We provide in this appendix four examples (Bright Star, Tycho, GCP, WISE) for which the MOCs are produced at order 6,7,8 and 9.

- Bright Star Catalog [V/50]: 9110 sources

| Order | Number of cells | Cell number increase | |
|---|---|---|---|
| 6 | 7939 | | 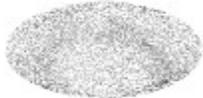 |
| 7 | 8630 | x 1.09 | 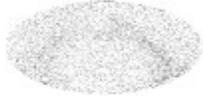 |
| 8 | 8842 | x 1.02 | 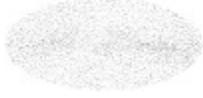 |
| 9 | 8934 | x 1.01 | 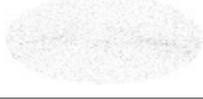 |

- General Catalog of Photometric Data [II/167]:165 896 sources

| Order | Number of cells | Cell number increase | |
|---|---|---|---|
| 6 | 19998 | | 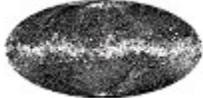 |
| 7 | 65616 | x 3.28 | 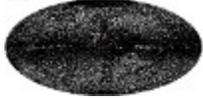 |
| 8 | 99879 | x 1.52 | 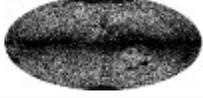 |
| 9 | 116950 | x 1,17 | 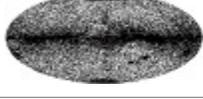 |

- Tycho [I/239]:1 058 332 sources

| Order | Number of cells | Cell number increase | |
|---|---|---|---|
| 6 | 783 | | 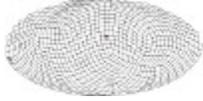 |
| 7 | 32320 | x 41.27 | 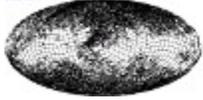 |
| 8 | 347064 | x 10.74 | 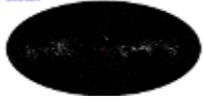 |
| 9 | 790483 | x 2.28 | 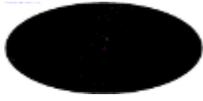 |

- WISE preliminary data catalog [II/307]: 257 310 278 sources

| Order | Number of cells | Cell number increase | |
|---|---|---|---|
| 6 | 1322 | | 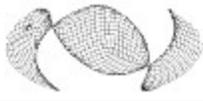 |
| 7 | 2845 | x 2.15 | 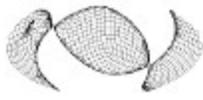 |
| 8 | 5709 | x 2 | 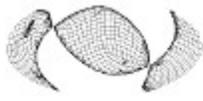 |
| 9 | 12231 | x 2.32 | 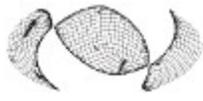 |